\documentclass[a4paper,11pt]{article}
\usepackage{pos}

\newcommand{\ksp}{\mathbf{k}}
\newcommand{\qedl}{\mathrm{QED}_{\mathrm{L}}}
\newcommand{\qedlir}{\mathrm{QED}_{\mathrm{L}}^{\mathrm{IR}}}
\newcommand{\qedr}{\mathrm{QED}_{\mathrm{r}}}

\newcommand{\qedc}{\mathrm{QED}_{\mathrm{C}}}

\title{Structure-dependent electromagnetic finite-volume effects through order $1/L^3$}
\ShortTitle{Structure-dependent electromagnetic finite-volume effects}

\author[a]{Matteo Di Carlo}
\author[a]{Maxwell T.~Hansen}
\author*[a,b]{Nils Hermansson-Truedsson}
\author[a]{Antonin Portelli}

\affiliation[a]{The Higgs Centre for Theoretical Physics, School of Physics and Astronomy, The University of Edinburgh\\
  Mayfield Rd, Edinburgh, EH9 3JZ, United Kingdom}

\affiliation[b]{Division of Particle and Nuclear Physics, Department of Physics, Lund University,\\
Professorsgatan 1, Box 118, 221 00, Lund, Sweden}

\emailAdd{nils.hermansson-truedsson@ed.ac.uk}

\abstract{We consider electromagnetic finite-volume effects through order $1/L^3$ in different formulations of QED, where $L$ is the periodicity of the spatial volume. An inherent problem at this order is the appearance of structure-dependent quantities related to form factors and the analytical structure of the correlation functions. The non-local constraint of the widely used $\qedl$ regularization gives rise to structure-dependent effects that are difficult to evaluate analytically and can act as a precision bottleneck in lattice calculations. For this reason, we consider general volume expansions relevant for the mass spectrum as well as leptonic decay rates in $\qedc$, $\qedl$ and $\qedlir$, the latter being a class of non-local formulations generalising $\qedl$. One choice within this class is $\qedr$, first introduced at this conference, and we show that the effects of non-locality for the $1/L^3$ term in the expansion can be removed. We observe that there are still $1/L^3$ contributions unrelated to the (non-)locality of the studied QED formulations, but rather to collinear singularities in the physical amplitudes.
}

\FullConference{The 40th International Symposium on Lattice Field Theory (Lattice 2023)\\
July 31st - August 4th, 2023\\
Fermi National Accelerator Laboratory\\}

\begin{document}
\maketitle

\section{Introduction}
\vspace{-0.4cm}
Lattice quantum chromodynamics (QCD) offers a systematically improvable approach for computing observables where non-perturbative dynamics play a crucial role. With relative precision in lattice calculations approaching the (sub-)percent level~\cite{DiCarlo:2019thl,Boyle:2022lsi}, one has to include effects from the up- and down-quark mass difference as well as electromagnetism. While the strong isospin-breaking from quark masses poses computational complications, including electromagnetic effects in the simulations is fundamentally more challenging~\cite{Hayakawa:2008an}. The underlying issue is constraining the long-range nature of the electromagnetic force into finite-volume spacetimes. Gauss' law in fact forbids charged states in finite volumes with periodic boundary conditions due to zero-momentum modes of photons, but there are several ways to circumvent the problem through dedicated finite-volume formulations of quantum electrodynamics (QED)~\cite{Hayakawa:2008an,Endres:2015gda,Lucini:2015hfa,Bussone:2017xkb,Davoudi:2018qpl}. Different prescriptions affect the volume-dependence of calculated observables in different ways. In this talk we analytically study finite-volume effects in formulations where the dependence scales as a polynomial of inverse powers of the spatial volume-extent $L$, along with potential logarithmic divergences $\log L$. 

It is very valuable to analytically subtract the leading volume scaling~\cite{BMW:2014pzb,Lubicz:2016xro,DiCarlo:2019thl,DiCarlo:2021apt,Boyle:2022lsi}. The isospin-breaking corrections to leptonic decays $P\rightarrow \ell \nu _\ell$, where $P$ is a pion or kaon and $\ell$ a lepton with corresponding neutrino $\nu _\ell$, have been calculated in Refs.~\cite{DiCarlo:2019thl,Boyle:2022lsi} in lattice QCD with electromagnetic corrections in the $\qedl$ formulation~\cite{Hayakawa:2008an}. It was recently observed that the $1/L^3$ correction in $\qedl$ potentially is very sizeable~\cite{DiCarlo:2021apt,Boyle:2022lsi}, thus motivating an analytical evaluation of the $1/L^3$ coefficient in the expansion. The dependence on the internal structure of the decaying meson in this coefficient can be handled using the method of Ref.~\cite{DiCarlo:2021apt} building on form factor decompositions and electromagnetic Ward identities. 

In this talk, we present a general study of the evaluation of structure-dependent finite-volume effects through order $1/L^3$ in QED formulations with power-law scaling. 
In particular we consider $\qedlir$, introduced in Ref.~\cite{Davoudi:2018qpl}. In this approach, the 
photon propagator is modified by reweighting some number of low momentum modes, generally to achieve some desired behaviour in the large-volume expansion of a given observable. $\qedl$ is a special case of this in which only the zero-momentum mode is modified by weighting it to zero, i.e. quenching it from the theory. $\qedr$ is another particular alternative designed to remove the $1/L^3$ terms from a variety of observables, first introduced at this conference~\cite{DiCarloLattice:2023} and discussed in more detail below. We apply the results to leptonic decay rates in the framework of $\qedlir$, and for the first time determine the $1/L^3$ contribution. We compare the values of 
 finite-volume coefficients in $\qedl$, $\qedr$ as well as $\qedc$~\cite{Lucini:2015hfa}. We observe that one finite-volume coefficient can lead to order $1/L^3$ effects in $\qedc$. 
\vspace{-0.3cm}

\section{Generalities on the finite-volume expansion}
\vspace{-0.4cm}
Let us consider an observable $\mathcal{O}$ in QCD+QED in a spacetime of geometry $V = \mathbb{R}\times L^3$. We will only be interested in leading-order QED corrections, 
i.e.~order $\alpha = e^2 /(4\pi)$ in the fine-structure constant $\alpha$. A determination of the observable in lattice simulations induces a volume-dependence, $\mathcal{O}(L)$, and the finite-volume effects are given by the difference $\Delta \mathcal{O}(L) =  \mathcal{O}(L) - \mathcal{O}(\infty)$. We restrict ourselves to the case where $\mathcal{O}(L)$ at most is logarithmically infrared divergent, as is relevant for leptonic decays. 
In QED formulations with a massless finite-volume photon propagator, $\Delta \mathcal{O}(L)$ will in addition to the infrared divergence $\kappa _{\log} \log L$ contain a series of terms $\kappa _n /L^n$. 
The coefficients $\mathcal{\kappa}_n$ depend on the chosen QED formulation, on the physical process and in general on internal structure of the interacting particles. The formulation dependence stems from the different ways of handling photon zero-momentum modes. In $\qedlir$ (in particular $\qedl$ and $\qedr$) the problematic modes are removed on each time-slice, rendering the theory non-local~\cite{Davoudi:2018qpl}. In $\qedc$, charge-conjugated boundary conditions are introduced, which in a local way forbids zero-momentum modes but introduces flavour mixing and charge non-conservation over the boundary~\cite{Lucini:2015hfa}.   

It is well-known that the $\kappa _n$ for the mass spectrum are independent of structure through order $1/L^2$~\cite{BMW:2014pzb,Lucini:2015hfa,Davoudi:2018qpl}, and for leptonic decays through order $1/L$~\cite{Lubicz:2016xro,DiCarlo:2021apt}.  The leading structure-dependence for these observables was determined for $\qedl$ in Ref.~\cite{DiCarlo:2021apt}, but it was observed that there are particular structure-dependent contributions at order $1/L^3$ due to the non-locality of $\qedl$. These contributions correspond to branch-cuts in the underlying correlation functions defining the physical process, e.g.~the Compton scattering tensor in the case of the mass spectrum~\cite{DiCarlo:2021apt}. 
\vspace{-0.3cm}
\section{Finite-volume expansion of the sunset topology}
\begin{figure}[t!]
	\centering
	\includegraphics[height=0.12\textheight]{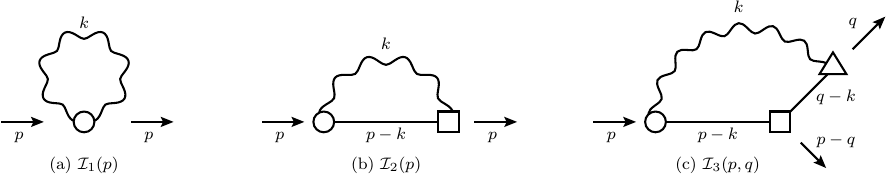}
	\caption{Three diagram topologies arising for the mass spectrum and leptonic decays, (a) the tadpole, (b) the sunset and (c) the triangle diagram. The momenta $p$ and $q$ are on-shell and correspond to massive particles, whereas $k$ is the virtual photon momentum. The circles, squares and triangles represent structure- and formulation-dependent vertex functions.}	\label{fig:masterdiagrams}
\end{figure}

\vspace{-0.4cm}
To obtain the volume-expansion one can use the method developed in Refs.~\cite{BMW:2014pzb,Lubicz:2016xro,DiCarlo:2021apt}, where the correlation function is skeleton expanded into a set of diagrams built from structure-dependent irreducible vertex functions. 
For our purposes we simply need to assume that such a decomposition has been made. Example topologies that appear for the mass spectrum and leptonic decays are shown in Fig.~\ref{fig:masterdiagrams}. In the following, we will be interested in finite-volume effects in the sunset diagram in Fig.~\ref{fig:masterdiagrams}(b), $\Delta \mathcal{I}_2$. 

The choice of QED formulation enters through the photon propagator for momentum $k=(k_0,\mathbf{k})$ and the set of allowed momenta $\mathbf{k}\in \Pi$. The sets are the same for $\qedlir$, $\qedl$ and $\qedr$, but different for $\qedc$ due to the charge-conjugated boundary conditions. Explicitly the sets are
\begin{align}\label{eq:pisets}
	& \qedlir : \quad \Pi = \Pi _{\textrm{L}}= \left\{ \left. \frac{2\pi \, \mathbf{n}}{L} \,  \right|  \mathbf{n}\in \mathbb{Z}^3 \setminus \{0,0,0\}  \right\}	\, ,
	\nonumber
	\\
	& \qedc : \quad \Pi = \Pi _{\textrm{C}} =  \left\{ \frac{2\pi }{L} \left(\left. \mathbf{n} + \frac{\bar{\mathbf{n}}}{2}\right)  \, \right|  \mathbf{n} \in \mathbb{Z}^3\, , \,  
	\bar{\mathbf{n}}=(1,1,1)
	\right\}	\, .
\end{align}
We write the Feynman-gauge propagator as
\begin{align}\label{eq:qedprop}
D_{\mu \nu} (k) = \delta _{\mu\nu }\, \frac{1+w_{|\mathbf{n}|^2}}{k_0^2+\mathbf{k}^2}\, , \qquad \mathbf{k}\in \Pi \, .
\end{align}
Here $w_{|\mathbf{n}|^2}$ is a discrete set of weights on photon-momentum modes $\mathbf{k}\in \Pi$, as was first introduced to define $\qedlir$~\cite{Davoudi:2018qpl}. However, we can use this form to consider also other formulations, in particular $w_{|\mathbf{n}|^2}=0$ for $\qedl$ and $\qedc$, and $w_{|\mathbf{n}|^2} = \delta _{\mathbf{n}^2,1}/6$ for $\qedr$. 

In this way the propagator $D_{\mu \nu} (k)$ in~(\ref{eq:qedprop}) allows for a common treatment of finite-volume effects of the QED formulations. For the sunset topology, we thus have
\begin{align}\label{eq:DeltaI2Def}
	\Delta \mathcal{I}_2
	=\left(  {\color{black} \frac{1}{L^3}\left. \sum _{\mathbf{\ksp} \in \Pi }\right.  - \int\frac{d ^3\ksp}{(2\pi)^3} } \right)  \int \frac{dk_{0}}{2\pi}\,  \frac{f \left( k_{0}, |\mathbf{k}| , \mathbf{p}\cdot  \mathbf{k} \right) }{ \left[ (p-k)^2+m^2\right]}\, \frac{ 1+ w_{|\mathbf{n}|^2}}{k^2} \, .
\end{align}
Here $m$ is the mass of the particle with momentum $p-k$, with $p^2 = -m^2$. We assume that the we are in a boosted frame with $p = (i\omega (\mathbf{p}),\mathbf{p})$. The numerator contains the observable-, structure and formulation-dependent function $f \left( k_{0}, |\mathbf{k}| , \mathbf{p}\cdot  \mathbf{k} \right) $, which is free of poles in $k_0$ but can contain branch-cuts. The arguments of $f$ show the possible ways the photon momentum can appear. 

The first step in obtaining a volume expansion in inverse powers of $L$ is to perform the $k_0$-integral in~(\ref{eq:DeltaI2Def}). As can be seen from the integrand, there are two poles in the upper half-plane, one from the photon propagator and one from the massive propagator. All the remaining analytic structure from potential branch-cuts resides in $f$ and we have
\begin{align}
	\int \frac{dk_{0}}{2\pi}\,  \frac{f \left( k_{0}, |\mathbf{k}| , \mathbf{p}\cdot  \mathbf{k} \right) }{ \left[ (p-k)^2+m^2\right]}\, \frac{1}{k^2}  = g^{\textrm{poles}}(\mathbf{k}) + g^{\textrm{rem}}(\mathbf{k})   \, .
\end{align}
Here $g^{\textrm{rem}}$ corresponds to the non-pole contributions. Consequently, the finite-volume effects are
\begin{align}
	\Delta \mathcal{I}_2 =\left(  {\color{black} \frac{1}{L^3}\left. \sum _{\mathbf{\ksp} \in \Pi }\right.  - \int\frac{d ^3\ksp}{(2\pi)^3} } \right)  \left( 1+ w_{|\mathbf{n}|^2} \right) \left[
		g^{\textrm{poles}}(\mathbf{k}) + g^{\textrm{rem}}(\mathbf{k})  \right]  = 	\Delta \mathcal{I}_2^{\textrm{poles}} + 	\Delta \mathcal{I}_2^{\textrm{rem}} \, .
\end{align}
From~(\ref{eq:pisets}) we see that the allowed momenta $\mathbf{k}$ in both $\Pi _{\mathrm{L}}$ and $\Pi _{\mathrm{C}}$ scale as $1/L$, which means that the integrand/summand can be expanded systematically in $L$. The coefficients in the expansion will contain derivatives of the function $f$ which depend
 on structure and the QED formulation, as well as a set of finite-volume coefficients defined by the sum-integral differences  
\begin{align}\label{eq:cjdef}
	{\color{black}{c}_{j}^w(\mathbf{v}) }= 	\left( \sum _{\color{black}\mathbf{n}\in \Pi} -\int d^3 \mathbf{n}\right) \frac{ 1+ w_{|\mathbf{n}|^2}  }{|{\mathbf{n}}|^j \, (1-\mathbf{v}\cdot \hat{{\mathbf{n}}})} \, . 
\end{align}
Here $\mathbf{v} = \mathbf{p}/\omega (\mathbf{p})$ is the velocity of the massive particle. Note that the factors $1/|\mathbf{n}|$ and $ 1/(1-\mathbf{v}\cdot \hat{{\mathbf{n}}})$ appear due to underlying soft and collinear singularities in the physical process. For a deeper discussion of the properties of these finite-volume coefficients, see Ref.~\cite{PortelliLattice:2023}. We also define $	{\color{black}{c}_{j}^w(\mathbf{0}) } = 	{\color{black}{c}_{j}^w}$.

The cut-contribution $g^{\textrm{rem}}(\mathbf{k})$ is by definition a smooth function at $|\mathbf{k}|= 0$. With $g^{\textrm{rem}}(\mathbf{0}) =  \mathcal{C}^{\textrm{rem}}$ the Poisson summation formula then immediately allows us to write (through order $1/L^3$)
\begin{align}\label{eq:i2remfves}
	\vspace{-0.5cm}
		\Delta \mathcal{I}_2^{\textrm{rem}} =  \frac{c_0^w \, \mathcal{C}^{\textrm{rem}}}{L^3}  \, . 
\end{align}
On the other hand, the pole contribution $g^{\textrm{poles}}(\mathbf{k})$ is not a smooth function, which in turn yields
\begin{align}\label{eq:i2polesfves}
		\Delta \mathcal{I}_2^{\textrm{poles}}&= 	\frac{f\, 
		{\color{black}{c}^w_2\left(\mathbf{v}\right)}}{16 \pi ^2 {\color{black}L} \omega (\mathbf{p})} 
	\nonumber \\	
	&
	+\frac{f^{(0,0,1)} \omega (\mathbf{p})
		\left({\color{black}{c}^w_1\left(\mathbf{v}\right)}-{\color{black}{c}^w_1}\right)+\left(f^{(0,1,0)}+i
		f^{(1,0,0)}\right)
		{\color{black}{c}^w_1\left(\mathbf{v}\right)}}{8 \pi  {\color{black}L^2} \omega (\mathbf{p})}
	\nonumber \\
	&
	+ \frac{1}{8{\color{black}L^3}}
	\Bigg\{ \frac{{\color{black}{c}^w_0\left(\mathbf{v}\right)}}{\omega (\mathbf{p})}
	\Big[
	f^{(0,2,0)} +\ldots
	\Big]
	-{\color{black}{c}^w_0}
	\Big[
	f^{(0,0,2)} \omega (\mathbf{p}) + \ldots
	\Big]
	\Bigg\} \, . 
\end{align}
In this expressions the function $f$ and its derivatives are evaluated at $(k_0=0,\, |\mathbf{k}| = 0,\, \mathbf{k}\cdot \mathbf{p} = 0)$. The ellipses contain other contributions from the function $f$, and have been introduced to avoid lengthy expressions. The crucial observation here is that branch-cut term $\mathcal{C}^{\textrm{rem}}$ only multiplies the coefficient $c_0^w$, while the poles contribute both to $c_0^w$ and $c_0^w(\mathbf{v})$. Also, we emphasise that both $c_j^w (\mathbf{v})$ and the function $f$ depend on the chosen QED prescription. Assuming that $f$ is known, the finite-volume effects for the sunset topology are given by~(\ref{eq:i2remfves})--(\ref{eq:i2polesfves}), where the coefficients $c_{j}^w(\mathbf{v})$ from~(\ref{eq:cjdef}) are defined by the allowed set of momenta in~(\ref{eq:pisets}) as well as the coefficients $w_{|\mathbf{n}|^2}$. We define $c_{j}^w(\mathbf{v}) = c_{j} (\mathbf{v})$ for $\qedl$, $c_{j}^w(\mathbf{v}) = \bar{c}_{j} (\mathbf{v})$ for $\qedr$ and $c_{j}^{w}(\mathbf{v}) = c_{j}^{\star} (\mathbf{v})$ for $\qedc$.       

As a final remark, the coefficient $c_0^w $ is non-zero in $\qedl$ due to the non-locality of the theory, $c_0=-1$. On the contrary, $c_{0}^\star = 0$ in $\qedc$, meaning that the branch-cut does not contribute in this case. The particular choice of parameters $w_{|\mathbf{n}|^2}$ in $\qedr$ gives $\bar{c}_0 = 0$, even though the theory has removed the zero-modes in a non-local fashion. This property makes $\qedr$ an interesting possible formulation to use in future lattice calculations, and we are currently performing a dedicated volume study in this framework to validate its suitability for future work~\cite{DiCarloLattice:2023}. In Section~\ref{sec:numerical} we numerically study the coefficients appearing in~(\ref{eq:i2remfves})--(\ref{eq:i2polesfves}) for $\qedl$, $\qedr$ and $\qedc$. 
\vspace{-0.4cm}

\section{Leptonic decay rates in $\qedlir$}
\vspace{-0.4cm}
We now apply the above formalism to derive the finite-volume effects for the leptonic decay $P\rightarrow \ell \nu _\ell$ through order $1/L^3$ in $\qedlir$. This extends the results in Refs.~\cite{Lubicz:2016xro,DiCarlo:2021apt,Boyle:2022lsi}, and requires expanding also the diagrams in Figs.~(a) and (c). Although we here limit the order to $1/L^3$ we have derived all structure-dependent vertex functions needed for leptonic decays, which would contribute also beyond order $1/L^3$. These will be presented in future work. 
Leaving out details about leptonic decays that can be found in Ref.~\cite{DiCarlo:2021apt}, the relevant function here is $Y^{(3)}(L)$ that in the rest frame of the decaying meson is found to be
\begin{align} \label{eq:y3}
	& {\color{black}Y^{(3)}(L) } = \ \frac34 + 4\, \log\left(\frac{m_\ell}{m_W}\right)  \ +\frac{{\color{black}c^{\textrm{IR}}_3}-2\, {\color{black}c^{\textrm{IR}}_3(\mathbf{v}_\ell)}}{2\pi}\,  - 2\,A_1(\mathbf{v}_\ell) + 2\,\log\left(\frac{m_W L}{4\pi}\right) \nonumber \\
	& 
	- 2\,A_1(\mathbf{v}_\ell)\left[\log\left(\frac{m_P L}{4\pi}\right)+\log\left(\frac{m_\ell L}{4\pi}\right)\right] 
	-\, \frac{1}{m_P L} \left[ \frac{(1+r_\ell^2)^2\,{\color{black}c^{\textrm{IR}}_2 } - 4\, r_\ell^2 \, {\color{black}c^{\textrm{IR}}_2(\mathbf{v}_\ell)}}{ 1-r_\ell^4} \right]
	\nonumber\\
	& + \, \frac{1}{(m_P L)^2} \Bigg[ - \frac{{\color{black}F_A^P}}{f_P} \, \frac{4\pi \, m_P \,\left[(1+r_\ell^2)^2 \, {\color{black}c^{\textrm{IR}}_1} - 4\, r_\ell^2\, {\color{black}c^{\textrm{IR}}_1(\mathbf{v}_\ell)}\right]}{1-r_\ell^4} 
	+ \frac{8\pi \, \left[(1+r_\ell^2) \, {\color{black}c^{\textrm{IR}}_1} - 2\,  {\color{black}c^{\textrm{IR}}_1(\mathbf{v}_\ell)}\right]}{ (1-r_\ell^4)} \Bigg] \nonumber
	\nonumber \\
	&	+ \frac{32\pi^2\, m_P}{ f_P  (1-r_\ell ^4)\, (m_PL)^3} \Bigg\{
	{\color{black} c^{\textrm{IR}}_0 (\mathbf{v}_\ell)} \, \left[ {\color{black}F_V^P }-{\color{black}F_A^P} + 2 m_P^2 r_\ell ^2 \, {\color{black}A^{(0,1)} \left(0,-m_P^2\right)} \right]
	+ {\color{black} c^{\textrm{IR}}_0}\, {\color{black} \mathcal{C}_\ell}  
	\Bigg\}	\, .
\end{align}
Here the $c_j^{\textrm{IR}}(\mathbf{v}_{\ell})$ are $\qedlir$ coefficients depending on the lepton velocity $\mathbf{v}_\ell$, $m_\ell$ the lepton mass, $m_W$ the $W$-boson mass, $m_P$ the meson mass, $r_\ell = m_\ell / m_P$ and $A_1(\mathbf{v}_\ell)$ a known function~\cite{DiCarlo:2021apt}. The branch-cuts of the correlation function appear in $\mathcal{C}_\ell$ together with structure-dependent form factors and point-like contributions. It is at the moment unknown how one would calculate the branch-cut. 
The coefficient $c_0^{\textrm{IR}}(\mathbf{v}_\ell)$ multiplies only structure-dependent form factors, $F_{V}^P$, $F_{A}^P = A(0,-m_P^2)$ and its derivative $A^{(0,1)}(0,-m_P^2)$ (where $A$ is a function of $k^2$ and $p\cdot k$). These are related to the real radiative leptonic decay $P\rightarrow \ell \nu _\ell \gamma$ and can be determined on the lattice~\cite{Desiderio:2020oej}. We emphasise that without knowledge of $\mathcal{C}_\ell$ the above expression cannot be used in practice. However, in $\qedr$ we have $\bar{c}_0 = 0$ which circumvents the problem and allows for full control through order $1/L^3$. 
\vspace{-0.4cm}

\section{Numerical comparison of finite-volume coefficients}\label{sec:numerical}
\vspace{-0.4cm}
We now perform a numerical calculation of the finite-volume coefficients $c_j^w(\mathbf{v})$ and $c_j^w$ appearing in~(\ref{eq:i2remfves})--(\ref{eq:i2polesfves}), for $\qedl$, $\qedr$ and $\qedc$.  
The velocity is chosen to be $\mathbf{v} = |\mathbf{v}|(1,1,1)/\sqrt{3}$, with $|\mathbf{v}| =  0.912401$ corresponding to the muon velocity in $K\rightarrow \mu \nu _\mu $. The values of the coefficients for $j=0,1,2$ are shown in Table~\ref{table:cjv}. We immediately see the non-locality of $\qedl$ from $c_0=-1$, the locality of $\qedc$ from $c_0^\star = 0$, and the effects of non-locality being removed from the choice of action parameters in $\qedr$ with $\bar{c}_0=0$. The velocity-dependent coefficients $c_0(\mathbf{v})$, $\bar{c}_0(\mathbf{v})$ and $c_0^\star (\mathbf{v})$ are in fact all non-zero. This shows that despite the locality of $\qedc$ there can be finite-volume effects at order $1/L^3$. An interesting observation is that if one were to do an angular averaging over all directions of the lepton velocity, the averaged finite-volume coefficients reduce to those at zero-velocity according to  $\int d\Omega \, c_j^w (\mathbf{v}) \propto c_j^w$~\cite{Davoudi:2018qpl}.  This means that if $c_0^w$ is zero in a formulation, an angular averaging would remove all volume-effects related also to $c_0^w(\mathbf{v})$~\cite{DiCarloLattice:2023,PortelliLattice:2023}.

	\begin{table}[t!]
	\centering
	\begin{tabular}{|c|c|c|c|c|c|c|}
		\hline
		$j$  & 
		{\color{black}$c_{j}(\mathbf{v})$} & {\color{black}$\bar{c}_{j}(\mathbf{v})$} & {\color{black}$c^\ast _{j}(\mathbf{v})$} & {\color{black}$c_{j}$} & {\color{black}$\bar{c}_{j}$} &{\color{black}$c^\ast _{j}$} \\
		\hline
		2  & 
		-16.3454 & -14.9613 &-3.20674 & -8.91363 & -7.91363 &-5.49014 
		\\
		1   & 
		-5.73018 & -4.3461 & 3.51224 & -2.8373 & -1.8373 &-0.80194 
		\\
		\color{black}0  &
		\color{black} -2.12369 & \color{black}-0.7396 &	\color{black} 3.69273 & 	\color{black} -1 &  {\color{black}0} &	\color{black} 0 
		\\
		\hline
	\end{tabular}
		\caption{Finite-volume coefficients ${c}_{j}\left(\mathbf{v}\right)$, $\bar{c}_j (\mathbf{v})$ and ${c}^\ast_{j}\left(\mathbf{v}\right)$, in $\qedl$, $\qedr$ and $\qedc$, respectively. }
		\label{table:cjv}
\end{table}

\vspace{-0.4cm}
\section{Conclusions}
\vspace{-0.4cm}
In this talk we have considered structure-dependent finite-volume effects decaying as inverse powers of $L$. The study is relevant for $\qedlir$, $\qedl$, $\qedc$ and $\qedr$, the last of which was first introduced at this conference. We derived formulae through order $1/L^3$, highlighting the impact of (non-)locality in the QED formulations. A particularly important finding was the possibility to eliminate the full $1/L^3$ contribution for leptonic decays in $\qedr$, which previously was found to be a precision bottleneck in the literature. With a dedicated volume-study for leptonic decays in $\qedr$, we are currently investigating the suitability of this formulation in future lattice calculations. 
\vspace{-0.4cm}
\section*{Acknowledgements}
\vspace{-0.4cm}

M.~D.~C., M.~T.~H., and A.~P.~are supported in part by UK STFC grant ST/P000630/1. Additionally M.~T.~H.~is supported by UKRI Future Leader Fellowship MR/T019956/1.   
A.~P.~received funding from the European Research Council (ERC) under the European Union's Horizon 2020 research and innovation programme under grant agreement No 757646 and additionally under grant agreement No 813942.
N.~H.-T.~is funded in part by the Swedish Research Council, project number 2021-06638, and in part by the UKRI, Engineering and Physical Sciences Research Council, grant number EP/X021971/1. N.~H.-T.~wishes to thank the Higgs Centre for Theoretical Physics at The University of Edinburgh for hosting him as a visitor when part of this work was done. 

\newpage
\bibliographystyle{JHEP}
\bibliography{refs}

\end{document}